\documentclass[twocolumn,english,showpacs,prb,floatfix]{revtex4}%
\usepackage{amsfonts}
\usepackage{amsmath}
\usepackage{amssymb}
\usepackage{graphicx}%
\begin{document}
\title{First-Principles Calculation of Mg(0001) Thin Films: Quantum Size Effect and
Adsorption of Atomic Hydrogen }
\author{Ping Zhang}
\affiliation{Institute of Applied Physics and Computational
Mathematics, Beijing 100088, P.R. China}

\begin{abstract}
We have carried out first-principles calculation of Mg(0001) free-standing
thin films to study the oscillatory quantum size effect exhibited in the
surface energy, work function, interlayer relaxation, and adsorption energy of
the atomic hydrogen adsorbate. The quantum well states have been shown. The
calculated energetics and interlayer relaxation of clean and H-adsorbed Mg
films are clearly featured by quantum oscillations as a function of the
thickness of the film, with oscillation period of about 8 monolayers,
consistent with recent experiments. The calculated quantum size effect in H
adsorption can be verified by observing the dependence of H coverage on the
thickness of Mg(0001) thin films gown on Si(111) or W(110) substrate which has
been experimentally accessible.

\end{abstract}
\maketitle

\section{Introduction}

When the thickness of thin metal films approaches the nanoscale, the
oscillatory quantum size effects (QSE) associated with electronic confinement
and interference will occur\cite{Pag,Tes,Mey,Kaw} due to the splitting of the
energy-level spectrum into subbands normal to the plane of the films.
Confinement of electrons often leads to strongly modified physical properties.
It has been shown that a change in film thickness by just one atomic layer can
result in property variations on the order of $1/N$, where $N$ is the
thickness of the film in terms of monolayers (ML). The oscillatory QSE have
long been clearly observed in ultrathin metal overlayers on metal
substrates\cite{Miller}, mostly involving noble metals. On the other hand,
thin metal films on semiconductors may be the basis for novel devices
utilizing quantum-well states (QWS). Thus recent systematic experimental and
theoretical investigation of the QSE has mainly been focused to Pb films
deposited on
Si(111)\cite{Chen1,Yeh,Su,Chen2,Chiang1,Chiang2,Chiang3,Dil,Xue1,Chiang4,Chiang5,Chiang6,Xue2,Zhang,Silva,Ogando}
substrate.

In this paper we present a detailed first-principles study of the electronic
structure and adsorption energetics of Mg(0001) free standing films. The QSE
in some other free-standing metal films such as Al\cite{Fei,Kie,Cir},
Li\cite{Boe}, and Pb\cite{Saa,Wei} have been theoretically reported in
previous references. The present study is directly motivated by the recent
experimental demonstration that highly perfect ultrathin epitaxial Mg(0001)
films can be grown on Si(111) substrate by low-temperature deposition and
annealing\cite{Aba}, and on W(110) substrate\cite{Sch2,Koi,Sch3}. Using
angle-resolved photoemission spectroscopy (ARPES) technique, Aballe \textit{et
al}.\cite{Aba} have showed that each time a QWS state falls in the wave-vector
and energy range of the upper branch of the Mg $sp$ band, a new peak is
visible in the photoemission spectrum, with the thickness interval between the
two sequential peaks being about $8$ ML. Schiller \textit{et al}.\cite{Sch2}
have extensively measured the electronic structure of magnesium from Mg(0001)
monolayers to bulk. More recently, Koitzsch \textit{et al}. and Schiller
\textit{et al}. have reported the spin-splitting effects in ultrathin Mg(0001)
films due to the coupling of the Mg(0001)/W(110) interface electronic
structure and the QWS states. It is expected that further work related with
the Mg(0001) thin films will be reported afterwards. From this aspect a
thorough theoretical investigation of the QSE in Mg(0001), including the
energetics and the interlayer relaxation, is necessary and will be helpful for
the future experimental reference. It should be mentioned that the theoretical
study of QSE on Mg(0001) was initiated by Feibelman\cite{Fei}.

The other object in this paper is to study the QSE character in the atomic
adsorption energetics, which has been neglected by most of the previous
studies. Since the adsorption property is closely characterized by the
chemical bonding between the adsorbate and the surface of the substrate, thus
when the substrate is ultra-thin, the QSE in the substrate will also influence
the behavior of the surface adsorption. Here as a case study we choose the
atomic hydrogen as the adsorbate on Mg(0001), since H is the most simple
element, also since the influence of atomic hydrogen adsorption on the surface
electronic structure of the metals has been extensively studied without
emphasis on QSE. Our results show that in the ultrathin Mg(0001) films, the
adsorption energy of atomic H displays a well-defined QSE.

This paper is organized as follows: In Sec. II, the \textit{ab initio} based
method and computational details is outlined. In Sec. III, the surface
properties of the Mg(0001) films, including the electronic structure, surface
energy, work function, and interlayer relaxation, as a function of the
thickness of the films, are presented and discussed. Also the properties of
adsorption of atomic hydrogen monolayer onto Mg(0001) surface is discussed in
detail by presenting the sensitivity of the adsorption energy to the thickness
of the Mg(0001) films. Finally, Sec. IV contains a summary of the work and our conclusion.

\section{Computational method}

The calculations were carried out using the Vienna \textit{ab initio}
simulation package\cite{Vasp} based on density-functional theory with
ultrasoft pseudopotentials\cite{Vand} and plane waves. In the present film
calculations, free-standing Mg(0001) films in periodic slab geometries were
employed. The periodic slabs are separated by a vacuum region equal to 20
\AA . In all the calculations below, a surface ($1\times1$) was employed for
the supercell slab. The Brillouin-zone integration was performed using
Monkhorse-Pack scheme\cite{Pack} with a $11\times11\times1$ $k$-point grid,
and the plane-wave energy cutoff was set $250$ eV. Furthermore, the
generalized gradient approximation (GGA) with PW-91 exchange-correlation
potential has been employed with all atomic configurations fully relaxed.
First the total energy of the bulk hcp Mg was calculated to obtain the bulk
lattice constants. The calculated $a$- and $c$-lattice parameters are $3.201$
\AA and $5.186$ \AA , comparable with experimental\cite{Amo} values of $3.21$
\AA and $5.213$ \AA , respectively. The use of larger $k$-point meshes did not
alter these values significantly. A Fermi broadening of 0.1 eV was chosen to
smear the occupation of the bands around $E_{F}$ by the Fermi-Dirac function.

\section{Results and discussion}

\textit{Band structure}.---
%TCIMACRO{\TeXButton{TeX field}{\begin{figure}[tbp]
%\begin{center}
%\includegraphics[width=0.6\linewidth]{fig1.eps}
%\end{center}
%\caption
%{GGA energy bands and density of electron states (right panel) of hcp bulk Mg.
%The dashed line denotes Fermi level.} \label{fig1}
%\end{figure}}}%
%BeginExpansion
\begin{figure}[tbp]
\begin{center}
\includegraphics[width=1.0\linewidth]{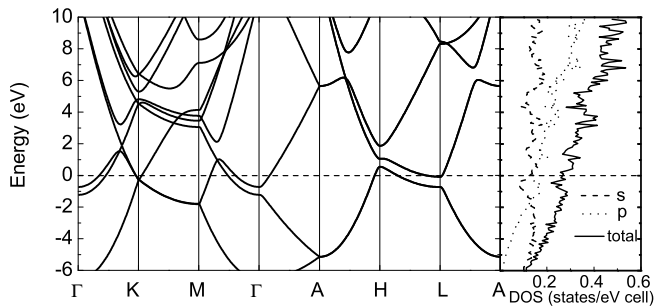}
\end{center}
\caption
{GGA energy bands and density of electron states (right panel) of hcp bulk Mg.
The dashed line denotes Fermi level.} \label{fig1}
\end{figure}%
%EndExpansion
We first studied the properties of electronic structures of Mg(0001) films. As
a first step, we present in Fig. 1 the band structure and the density of
states (DOS) of bulk hcp Mg. One can see that the DOS of bulk Mg is nearly
free-electron like ($\simeq\sqrt{\epsilon}$). This is different from its close
neighbor Be whose DOS resembles somewhat that of a semiconductor due to the
absence of core electrons in Be atoms. Also one can see From Fig. 1 that there
are two filled state at $\Gamma$ with energies around $1$ eV, while in the
case of Be the corresponding states are above the Fermi energy. Although the
outmost electronic configuration of elemental Mg is $3s^{2}$, one can see from
Fig. 1 that the $p$-orbital component in bulk Mg plays as well an important
role around $E_{F}$.%

%TCIMACRO{\TeXButton{TeX field}{\begin{figure}[tbp]
%\begin{center}
%\includegraphics[width=0.6\linewidth]{fig2.eps}
%\end{center}
%\caption{Calculated (GGA) energies at $\Gamma
%$ in Mg(0001) thin films as a function of
%thickness, with the energy set to zero at the Fermi level. The right
%panel replot the bulk energy dispersion in the [0001] direction.} \label{fig2}
%\end{figure}}}%
%BeginExpansion
\begin{figure}[tbp]
\begin{center}
\includegraphics[width=1.0\linewidth]{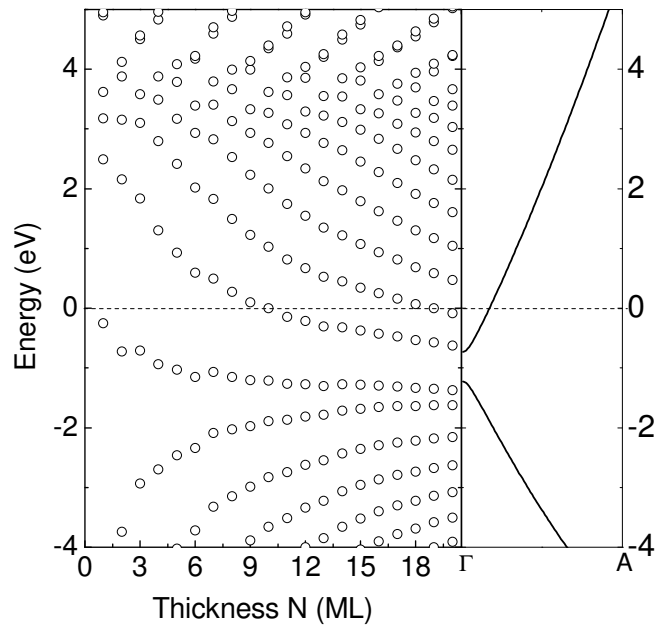}
\end{center}
\caption{Calculated (GGA) energies at $\Gamma
$ in Mg(0001) thin films as a function of
thickness, with the energy set to zero at the Fermi level. The right
panel replot the bulk energy dispersion in the [0001] direction.} \label{fig2}
\end{figure}%
%EndExpansion
The electronic structure properties of Mg(0001) film are shown in Fig. 2 as a
function of the thickness of the film. Here we only plot the energy levels at
$\overline{\Gamma}$ point without interlayer relaxation. The energy zero is
set at the Fermi level of each film. Interlayer relaxation effect has also
been studied and it is found that the overall thickness dependence of the
energies is similar to that without relaxation. For comparison and
illustration, the bulk energy dispersion along the [0001] ($\Gamma\rightarrow
A$) direction [see Fig. 1], which determines the energy range for the quantum
well states, is plotted again in Fig. 2 (right panel). One can see that the
energy gap at $\Gamma$ is small due to the hybridization of Mg $s$ and $p$
orbitals, with the energy of $\sim$0.5 eV. The QWS states arise from the upper
band. When the thickness of the film is increased to $\sim$8 ML, then a QWS
state, with the energy crossing the Fermi level, occurs. The next energy
crossing with the Fermi level occurs at the film thickness of 16 ML. Our
calculated results of QWS states are in good agreement with recent
experimental ARPES measurement.

For $sp$ metals the QWS states are often analyzed in the framework of the
phase accumulation model\cite{Ech,Smith}. Here the free-standing Mg(0001) film
is considered as a quantum well confining electrons between the two vacuums in
the slab. Only such $k_{\bot}$ (perpendicular component of bulk wave vector)
values of the electrons are allowed that fulfill the stationary state
condition for integer $n$,%
\begin{equation}
2k_{\bot}Nd+2\Phi=2\pi n,\tag{1}%
\end{equation}
where $N$ is the number of atomic layers in the film, $d=c/2$ the interlayer
spacing, and $\Phi$ the phase shift of the electronic wave function upon
reflection at the film-vacuum interface. Using Eq.(1) one can calculate the
periodicity for the QWS states crossing the Fermi level, $\Delta
N=\pi/(k_{\bot}^{f}d)$, where $k_{\bot}^{f}$ is the perpendicular component of
Fermi wave vector. From the right panel in Fig. 2, one can see that the upper
branch of the bulk $sp$ band runs through about $25\%$ of the Brillouin zone,
$k_{\bot}^{f}=0.25\pi/c$. One gets $\Delta N=8$. Therefore a new QWS state
occurs every $8$ ML, which is verified in the left panel in Fig. 2 that an
energy branch moves down, crossing the Fermi level for every incremental
increase in the film thickness of $8$ ML.

\textit{Energetics}.--- Figure 3(a) shows the total energy $E_{t}(N)/N$ per ML
as a function of the thickness of the Mg(0001) film. The atoms in the slabs
have been fully relaxed during calculations. One can see from Fig. 3(a) that
with increasing the thickness, $E_{t}(N)/N$ gradually approaches a constant
value which in the limit is equal to the energy per atom in the bulk Mg.%

%TCIMACRO{\TeXButton{TeX field}{\begin{figure}[tbp]
%\begin{center}
%\includegraphics[width=0.6\linewidth]{fig3.eps}
%\end{center}
%\caption{(a) Monolayer energy $E(N)/N$, (b)
%corresponding energy difference $\Delta
%E(N)$, and (c) surface energy for fully
%relaxed Mg(0001) $1\times1$ slabs as
%a function of thickness.} \label{fig3}
%\end{figure}}}%
%BeginExpansion
\begin{figure}[tbp]
\begin{center}
\includegraphics[width=1.0\linewidth]{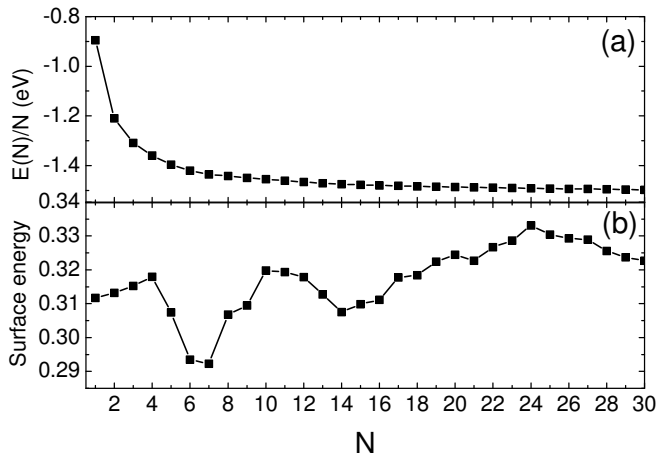}
\end{center}
\caption{(a) Monolayer energy $E(N)/N$, (b)
corresponding energy difference $\Delta
E(N)$, and (c) surface energy for fully
relaxed Mg(0001) $1\times1$ slabs as
a function of thickness.} \label{fig3}
\end{figure}%
%EndExpansion
An energetic quantity more suitably tailored to the QSE is the surface energy
which is defined as one-half of the energy difference between the film and the
bulk with the same number of atoms, including the proper subtraction of a term
linear in $N$\cite{Boe2}. The thickness dependence of surface energy is shown
in Fig. 3(b). It reveals that consistent with the result of the electronic
structure in Fig. 2, the surface energy follows a simple oscillatory form with
the period of $\Delta N\simeq8$. These oscillations arise from the occupation
of electronic levels close to Fermi surface. Also one can see that the
oscillation pattern is not as good as one expects. This is due to the fact
that the atomic arrangement in the slabs has been fully relaxed during the calculation.%

%TCIMACRO{\TeXButton{TeX field}{\begin{figure}[tbp]
%\begin{center}
%\includegraphics[width=0.6\linewidth]{fig4.eps}
%\end{center}
%\caption{Work function of clean and H-adsorbed Mg(0001)
%thin films as a function of thickness.} \label{fig4}
%\end{figure}}}%
%BeginExpansion
\begin{figure}[tbp]
\begin{center}
\includegraphics[width=1.0\linewidth]{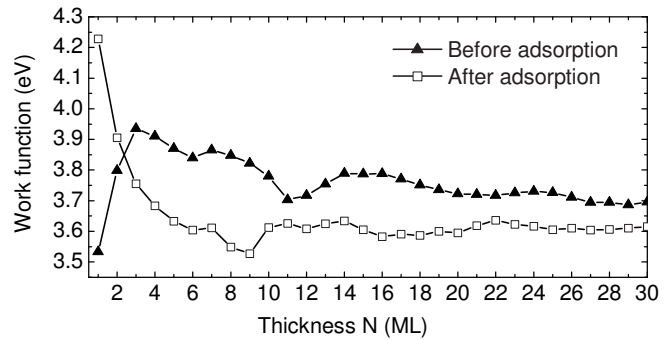}
\end{center}
\caption{Work function of clean and H-adsorbed Mg(0001)
thin films as a function of thickness.} \label{fig4}
\end{figure}%
%EndExpansion
Figure 4 (line with triangles) shows the work function as a function of the
thickness of Mg(0001) film for relaxed atomic geometry. One can see the work
function is also featured by an oscillatory form multiplied by a damping
factor. The oscillation period in thickness is again $\Delta N\simeq8$.

\begin{table}[th]
\caption{Interlayer relaxations given in percent, $\Delta d_{i,i+1}$, of
Mg(0001) as a function of the thickness of the film.}
%\label{specs}%
\begin{tabular}
[c]{ccccccc}\hline\hline
{$N$} & $\Delta d_{12}$ & $\Delta d_{23}$ & $\Delta d_{34}$ & $\Delta d_{45}$
& $\Delta d_{56}$ & $\Delta d_{67}$\\\hline
2 & +8.387 &  &  &  &  & \\
3 & +6.985 & +6.995 &  &  &  & \\
4 & +1.766 & -2.104 & +1.779 &  &  & \\
5 & +1.593 & -0.672 & -0.67 & +1.595 &  & \\
6 & +0.785 & -1.161 & +0.604 & -1.169 & +0.798 & \\
7 & +2.787 & -0.282 & -0.051 & -0.1 & -0.237 & +2.817\\
8 & +2.323 & -0.741 & -0.605 & -0.1 & -0.652 & -0.7\\
9 & +1.603 & -0.903 & -0.192 & -0.46 & -0.46 & -0.187\\
10 & +1.438 & -1.075 & -0.053 & -0.341 & -0.452 & -0.342\\
11 & +0.654 & -1.1 & +0.066 & -0.213 & -0.34 & -0.34\\
12 & +0.537 & -1.089 & +0.212 & -0.583 & +0.059 & -0.533\\
13 & +0.898 & -1.304 & 0.227 & -0.358 & -0.375 & -0.105\\
14 & +1.172 & -1.197 & 0.115 & -0.412 & -0.236 & -0.391\\
15 & +1.723 & -0.93 & -0.264 & -0.194 & -0.468 & -0.208\\\hline\hline
\end{tabular}
\end{table}

\textit{Interlayer relaxation}.---
%TCIMACRO{\TeXButton{TeX field}{\begin{figure}[tbp]
%\begin{center}
%\includegraphics[width=0.6\linewidth]{fig5.eps}
%\end{center}
%\caption{Interlayer relaxations of Mg(0001) thin films as a function
%of thickness.} \label{fig5}
%\end{figure}}}%
%BeginExpansion
\begin{figure}[tbp]
\begin{center}
\includegraphics[width=1.0\linewidth]{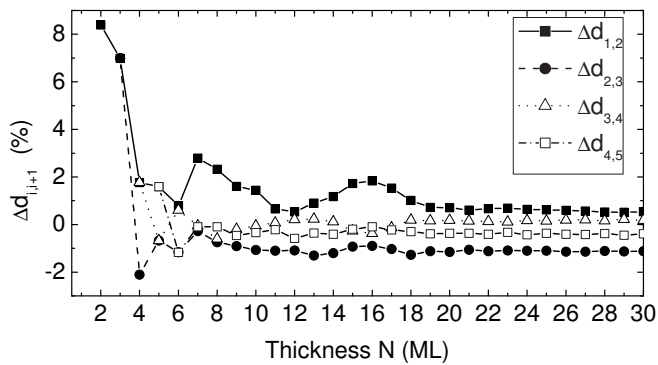}
\end{center}
\caption{Interlayer relaxations of Mg(0001) thin films as a function
of thickness.} \label{fig5}
\end{figure}%
%EndExpansion
The interlayer relaxation, $\Delta d_{i,i+1}$, is given in percent with
respect to the unrelaxed interlayer spacing, $d_{0}$, i.e., $\Delta
d_{i,i+1}=100(d_{i,i+1}-d_{0})/d_{0}$. $d_{i,i+1}$ is the interlayer distance
between two adjacent layers parallel to the surface calculated by total energy
minimization. $d_{0}=c/2$ is the bulk interlayer distance. As mentioned above,
all layers in the slab were allowed to relax. Obviously, the signs $+$ and $-$
of $\Delta d_{i,i+1}$ indicate expansion and contraction of the interlayer
spacings, respectively. The relaxation of Mg(0001) surface as a function of
the thickness of the film is summarized in Table I. Furthermore, the
interlayer relaxations are also plotted in Fig. 5 for more clear illustration.
One can see: (i) The two outmost layers relax significantly from the bulk
value, in agreement with experimental observation\cite{Ismail}. In the whole
range of layers that we considered, the topmost interlayer relaxation is
always outward ($\Delta d_{1,2}>0$). The value of $\Delta d_{1,2}$ starts from
+8\% for a slab with only 2 ML, and approaches a final value of $\sim$1\% with
increasing the thickness of Mg(0001) film. In contrast to the behavior of
$\Delta d_{1,2}$, the second interlayer relaxation is always inward ($\Delta
d_{2,3}<0$). Note that the first interlayer separation on most metal surfaces
is contracted, Mg(0001) is one of the few exceptions; (ii) The interlayer
spacings oscillate as a function of the thickness of the film with a damped
magnitude. The oscillation period is about $8$ ML in the thickness, thus
clearly indicating the QSE in the interlayer relaxation. After 30 ML, which is
the maximal layers considered here, the oscillations are invisible, which
suggests that the semi-infinite surface limit is now reached.

\textit{Adsorption of atomic hydrogen: QSE of binding energy}.--- To further
illustrate the physical properties influenced by finite size of the thin
films, in this section we focus our attention to the adsorption of atomic
hydrogen on Mg(0001) thin films. To the best of our knowledge, the reflection
of QSE by the adsorption features has been neglected by most of the previous studies.%

%TCIMACRO{\TeXButton{TeX field}{\begin{figure}[tbp]
%\begin{center}
%\includegraphics[width=0.4\linewidth]{fig6.eps}
%\end{center}
%\caption{The four different adsorption sites for H adatom on
%Mg(0001) surface.
%} \label{fig6}
%\end{figure}}}%
%BeginExpansion
\begin{figure}[tbp]
\begin{center}
\includegraphics[width=0.5\linewidth]{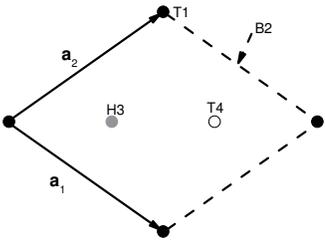}
\end{center}
\caption{The four different adsorption sites for H adatom on
Mg(0001) surface.
} \label{fig6}
\end{figure}%
%EndExpansion
Before we study the hydrogen adsorption properties as a function of the
thickness of the film, we need to determine the energetically favorable
adsorption site. Since the preference of adsorption site is not sensitive to
the thickness of the substrate, thus to look for this preference, it is
sufficient to give a study on the slabs with fixed thickness of Mg(0001)
substrate, which at present is chosen to be 9 ML. We choose four most probable
adsorption sites, namely, on-top (T1), bridge (B2), fcc (H3), and hcp (T4)
sites, which are schematically indicated in Fig. 6. The binding energy is
calculated using the following equation: Binding energy [atomic H]$=-$%
($E[$H/Mg(0001)$]-E[$Mg$(0001)]-2E[$H$]$)/2 where $E[$H/Mg(0001)$]$ is the
total energy of a slab which consists of 9 layers of Mg atoms and one H atom
on each side keeping inversion symmetry, $E[$Mg$(0001)]$ is total energy of
the slab without H atoms, and $E[$H$]$ is total energy of a free H atom which
is put in a supercell with the size of 10 \r{A}. As a result, the calculated
binding energy of atomic H for different adsorption configurations is 1.317 eV
(T1), 2.394 eV (H3), and 2.319 eV (T4). The B2 site is unstable and after the
zero-temperature relaxation, the H adatom at B2 will diffuse to H3 site. Thus
the fcc (H3) site is most stable and in the following discussions, the atomic
H is always put on H3 site during the simulation. Note that this H3 preference
for H adsorption on Mg(0001) is different from the cases in Be(0001) and
W(100) wherein the bridge-site is preferred\cite{Rol,Bar}.%

%TCIMACRO{\TeXButton{TeX field}{\begin{figure}[tbp]
%\begin{center}
%\includegraphics[width=0.6\linewidth]{fig7.eps}
%\end{center}
%\caption{Calculated (a) binding energy of H adatom and (b) adsorbate
%height as a function thickness of Mg(0001) films.} \label{fig7}
%\end{figure}}}%
%BeginExpansion
\begin{figure}[tbp]
\begin{center}
\includegraphics[width=1.0\linewidth]{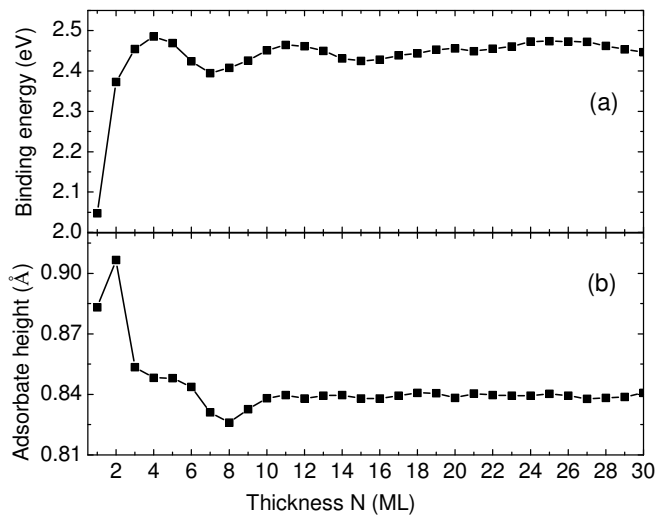}
\end{center}
\caption{Calculated (a) binding energy of H adatom and (b) adsorbate
height as a function thickness of Mg(0001) films.} \label{fig7}
\end{figure}%
%EndExpansion
After finding the preferred atomic H adsorption site (H3), we gave a series of
calculations for the binding energy of the H adsorbate as a function of the
thickness of Mg(0001) thin films. The results are summarized in Fig. 7(a). One
can see that the binding energy curves up at small film thickness, followed by
damped oscillations when increasing the Mg ML in the slab. Thus the binding
energy of atomic H depends on the thickness of the quantum films in an
oscillatory way. The oscillation period in thickness is about $8$ monolayers,
indicating a well-defined QSE in the adsorption of atomic H on Mg(0001). In
experiment this QSE of atomic adsorption can be observed by investigating the
dependence of H coverage on the monolayers of Mg(0001) thin films. Also we
have calculated the H adsorbate height and the results are plotted in Fig.
7(b), which again shows the periodic oscillations indicative of QSE.
Furthermore, we have calculated the work function for the H-adsorbed Mg(0001),
which is shown in Fig. 4 (line with squares). One can see that compared to the
clean Mg(0001), the work function is reduced by the presence of H adlayer,
implying that the charge is transferred from H to Mg. Also it shows in Fig. 4
that as in the case of clean surface, the work function of H-adsorbed Mg(0001)
is oscillatory in the amplitude with respect to the thickness of the film. The
oscillation period in thickness is again $\Delta N\simeq8$.

\section{Conclusion}

In summary, the Mg(0001) thin films have been studied by density-functional
theory pseudopotential plane-wave calculations. The dependence of electronic
structure, energetics, and interlayer relaxation upon the thickness of the
film has been fully investigated, clearly showing the metallic QSE of the
film, in consistent with the recent experiments. We have also studied the
atomic hydrogen adsorption on the Mg(0001) film. It has been shown that the
adsorption energy of H oscillates with the increase of Mg monolayers in the
slab. As the other energetic quantities, this oscillation in the adsorption
energy can also be explained by the occurrence of the QWS states in the
ultrathin metal films.

\begin{acknowledgments}
This work was supported by the CNSF No. 10544004 and 10604010.
\end{acknowledgments}

\end{document}